# Aggregate formation of surface-modified nanoparticles in solvents and polymer nanocomposites

Dafne Musino[1], Anne-Caroline Genix[1]*, Thomas Chaussée[2], Laurent Guy[2], Natalia Meissner[3], Radoslaw Kozak[3], Thomas Bizien[4], Julian Oberdisse[1]

[1] *Laboratoire Charles Coulomb (L2C), Université de Montpellier, CNRS, F-34095 Montpellier, France*

[2] *Solvay Silica, 15 rue Pierre Pays BP52, 69660 Collonges au Mont d'Or, France*

[3] *Synthos Spółka Akcyjna, Chemików 1, 32600 Oświęcim, Poland*

[4] *SOLEIL Synchrotron, L'Orme des Merisiers, Gif-Sur-Yvette, 91192 Saint-Aubin, France*

*\* Author for correspondence: anne-caroline.genix@umontpellier.fr*

**Abstract**

A new method based on the combination of small-angle scattering, reverse Monte Carlo simulations, and an aggregate recognition algorithm is proposed to characterize the structure of nanoparticle suspensions in solvents and polymer nanocomposites, allowing detailed studies of the impact of different nanoparticle surface modifications. Experimental small-angle scattering is reproduced using simulated annealing of configurations of polydisperse particles in a simulation box compatible with the lowest experimental q-vector. Then, properties of interest like aggregation states are extracted from these configurations, and averaged. This approach has been applied to silane surface-modified silica nanoparticles with different grafting groups, in solvents and after casting into polymer matrices. It is shown that the chemistry of the silane function – in particular mono- or trifunctionality possibly related to patch formation – affects the dispersion state in a given medium, in spite of an unchanged alkyl chain length. Our approach may be applied to study any dispersion or aggregation state of nanoparticles. Concerning nanocomposites, the method has potential impact on the design of new formulations allowing controlled tuning of nanoparticle dispersion.

# INTRODUCTION

The formation of nanoparticle (NP) aggregates in suspension or polymer matrices is directly related to the interactions between NPs. [1] Well-stabilized suspensions, e.g., are characterized by dominant repulsive interactions, which may be of electrostatic or steric origin. [2] On the other extreme, loss of colloidal stability induces the growth of aggregates, which may ultimately lead to phase separation. In many applications, the controlled formation of aggregates is desired, as such large-scale structures may serve as volume-spanning entities, allowing the optimization of transport properties, like electrical conductivity [3, 4], or mechanical strength [5]. The improvement of the latter property is known as mechanical reinforcement in polymer nanocomposites. [6-8] It depends on the volume spanning properties of aggregates, and thus on their internal structure. This effect comprises the increase in volume of aggregates with respect to their pure NP content due to bound or occluded rubber inside aggregates, which can be expressed as an increase of the global volume fraction of hard objects. Hydrodynamic reinforcement as usually described by the Einstein equation for viscosity adapted by Smallwood and others to moduli [9, 10] is thus found to increase. Furthermore, at higher particle content, the system approaches percolation more quickly if the structure of aggregates is more volume-spanning . [8, 11]

Small-angle scattering is a powerful tool for the analysis of multi-scale structures, [12, 13] as given by aggregation of nanoparticles into higher order structures. Unfortunately, the "inversion" of scattered small-angle intensities into real-space information like distribution- or correlation functions is an ill-posed problem, due to the loss of phase information during the intensity measurement, among others. Important conceptual progress has been made by Glatter and coworkers since the 1970s [14, 15], who proposed the indirect Fourier transform (IFT) method in order to extract pair distance distribution functions. IFT has been applied to many different soft condensed matter objects, like for example to polymeric micellar systems, [16, 17] NPs, [18-20] or complexes like protein-surfactant assemblies [21]. Note that the determination of objective solutions to the phase problem has been discussed recently. [22] The Glatter group has also included particle interactions in the generalized IFT (GIFT) method, [23-25] which has also been applied to systems as varied as emulsions [26] or surfactant micelles, [27, 28] including reverse micelles [29]. The strategy behind GIFT is to coherently describe particle interactions and internal structure by IFT. Since the early 1990s, a second, stochastic inversion method has been proposed, reverse Monte Carlo (RMC). [30-35] Originally designed for elemental liquids [36], the technique has been applied to the inversion of scattering or diffraction data of many disordered systems, including semiconductors, [37, 38] liquid water,[39] molecular liquids [40-42], ionic solutions [43-45] , or glasses [46]. In parallel, Svergun and coworkers have successfully applied similar approaches to the shape of biological molecules determined from small-angle scattering. [47-49] RMC has only rarely been used for small-angle scattering of self-assembled structures [50] or colloids [51], and in particular nanoparticle aggregation has only been studied from a simulation point of view [52-54]. The method has been adapted to aggregates in nanocomposites by one of us [55], while others solved the two-dimensional scattering problem of stretched nanocomposites by RMC [56]. In RMC, the physical entities (ions, nanoparticles) are used to construct real space solutions compatible with the data, whereas in IFT the structure is described in a basis of mathematical functions (typically cubic splines) which usually do not correspond to physical objects. On the other hand, IFT produces a single average correlation function, whereas RMC proposes a series of compatible configurations, which can then be studied and averaged. The latter technique is used here to generate polydisperse particle configurations, which will be analyzed in a second step by an aggregate recognition algorithm.

Interactions between NPs and thus their aggregation can be controlled via NP surface modification. [57, 58] Bare oxide NPs are covered with –OH groups and usually carry electrostatic charges in water, contributing to their electrostatic stabilization in this solvent, as opposed to attractive short-ranged Van der Waals forces. [59] Modifications of the interface by adsorption or chemical grafting may change the number of charges, [60] or the hydrophilic/hydrophobic character of the surface, e.g., with amphiphilic molecules [61-64]. Moreover, the formation of "patches", which may be surface micelles or adsorbed proteins was found to induce strong interactions. [65-67] Depending on the solvophilicity of such patches, the net interaction may be sterical repulsion, or attractive bridging. In the present article, the grafting of silane molecules having different reactive groups is used to modify interactions between NPs in precursor suspensions, and between NPs and a polymer matrix in nanocomposites. The resulting aggregation is studied by small-angle X-ray scattering. The scattering analysis of interacting particle systems is often focused on the description of the interaction peak, and the corresponding local interparticle distances. [68, 69] However, depending on the type of surface treatment, the interactions in solvents and polymer differ – which is precisely the desired effect of surface modification – and may induce aggregation, which is rarely described due to complicated correlations between possibly polydisperse aggregates. Moreover, in presence of particle polydispersity, superimposed to a distribution of aggregate masses, many differently weighted partial structure factors have to be determined and included in the quantitative data treatment. Needless to say, this is a hopeless endeavor in most practical cases.

We propose a combination of small-angle scattering of X-rays, reverse Monte Carlo analysis, and aggregate recognition to extract detailed information on colloidal aggregate mass distribution functions in NP suspensions and polymer nanocomposites. In particular, it is shown in this article that it is possible to describe all interaction terms between particles making up aggregates by focusing on particle configurations in real space, followed by the analysis of the latter in terms of aggregate distribution functions. Isolating aggregates in the configurations thus allows ignoring interactions. This will be shown to be decisive for the investigation of the influence of surface modification using silane molecules with different grafting functions, and is hoped to be useful for many other systems.

**METHODS**

**Nanoparticle suspensions and surface modification.** Colloidal silica NPs (Ludox® TM40, 40%w suspension in water) were purchased from Sigma-Aldrich. They have been characterized by SAXS in de-ionized water at 1%v ($R_0$ = 12.5 nm, log-normal polydispersity σ = 0.12, $V_{si}$ = 8728.7 nm$^3$). Different graftable silane molecules have been used for surface modification of the NPs by varying either the number (tri or mono) or the type (ethoxy or methoxy) of the functional group: octyl triethoxysilane termed $C_8$, octyl-trimethoxysilane ($C_{8m}$), and octyl-methoxy(dimethyl)silane ($C_{8mm}$), all from Sigma-Aldrich. The protocol consists in performing the grafting reaction over 24h in an ethanol-water mixture (63%v ethanol), at pH 9 and T = 50°C. The surface-modified silica NPs were characterized after washing cycles by centrifugation and redispersion in ethanol. The silane grafting density on NPs was measured by thermogravimetric analysis (TGA, Mettler Toledo, 5 K/min from 35°C to 900°C under air, 60 ml/min), and are given in Table 1. For the trialkoxysilanes ($C_8$ and $C_{8m}$), we assumed the reaction of two alkoxy groups with the hydroxyl groups on the silica surface. The suspensions are then transferred into pure ethanol and into MEK by dialysis (24h for each step, MEK twice), and sonicated for 30 minutes at room temperature. The remaining quantities of ethanol and water have been determined by NMR and were found to be below 2.5%. Comparative TGA experiments on the reaction suspension in ethanol-water after washing, and after dialysis into MEK have shown that (70 ± 7)% of the molecules added for the reaction are effectively grafted on the NPs, and that these are the only remaining both in MEK and in the final polymer nanocomposites.

**Table 1:** Silane molecules used for the surface modification of silica nanoparticles, and grafting density determined by TGA.

| Coating agent | Abbreviation | Linear formula | ρ (silane.nm$^{-2}$) |
|---|---|---|---|
| Octyl-triethoxysilane | $C_8$ | $C_{14}H_{32}O_3Si$ | 1.1 |
| Octyl-trimethoxysilane | $C_{8m}$ | $C_{11}H_{26}O_3Si$ | 1.2 |
| Octyl-methoxy(dimethyl)silane | $C_{8mm}$ | $C_{11}H_{26}OSi$ | 1.3 |

**Polymer and nanocomposite formulation.** Highly monodisperse non end-functionalized styrene-butadiene (SB; styrene units 19.1%w, butadiene 80.9%w, out of which 42.6%w 1,4-units, the rest 1,2) random copolymer was purpose-synthesized by Synthos (177 kg/mol, polydispersity index = 1.02). The polymer is dissolved in MEK (10%v), then mixed with the (previously surface-modified) NP suspension in MEK at 1%v (1h30 stirring), followed by solvent casting on a Teflon support for 24h at 50°C and drying for 24h under vacuum at room temperature.

**Structural analysis.** Small-angle X-ray scattering (SAXS) was performed on beamline SWING at synchrotron SOLEIL (Saint Aubin, France) using standard conditions (sample-to-detector distances 2 m, 5 m and 6.5 m; wavelength 1 Å, giving a q-range from 6.2 $10^{-4}$ Å$^{-1}$ to 5.6 $10^{-1}$ Å$^{-1}$). Standard data reduction tools given by Soleil were used (Foxtrot 3.1). Matrix contributions have been measured independently, and subtracted. For comparison of particle scattering in different solvents or polymer, at different particle concentrations, intensities have been normalized to the form factor scattering observed at 1%v in water. The normalized intensity $I_{norm}$ then reads:

$$I_{norm}(q) = I_{exp}(q) \frac{\Phi_{form}}{\Phi} \left(\frac{\Delta\rho_{form}}{\Delta\rho}\right)^2 \quad (1)$$

where the index 'form' corresponds to the form factor measurement, and $\Phi$ and $\Delta\rho$ to the particle volume fraction and contrast of the originally measured intensity $I_{exp}$. The scattering length densities used to calculate the contrasts $\Delta\rho = \rho_{SiO2} - \rho_{medium}$ are: $\rho_{SiO2}$ = 18.9 $10^{10}$ cm$^{-2}$, $\rho_{H2O}$ = 9.5 $10^{10}$ cm$^{-2}$, $\rho_{EtOH}$ = 7.6 $10^{10}$ cm$^{-2}$, $\rho_{MEK}$ = 7.7 $10^{10}$ cm$^{-2}$, and $\rho_{SB}$ = 8.9 $10^{10}$ cm$^{-2}$. The difference between the experimentally measured (and renormed) intensity and the one calculated by RMC was expressed via the difference of the structures factors as defined below, using the following definition of $\chi^2$:

$$\chi^2 = \frac{1}{N_q}\sum_{p=1}^{N_q} \left(\frac{S_{exp}(q_p) - S(q_p)}{\Delta S}\right)^2 \quad (2)$$

where $N_q$ is the number of data points (including all directions), and $\Delta S$ is the estimated experimental error of the structure factor, here set to 0.01 for all q values. This value corresponds typically to the error bar given in the synchrotron experiment over the relevant q-range.

**RESULTS**

**Reverse Monte Carlo and aggregate recognition.** The main idea of reverse Monte Carlo applied to colloids is to produce configurations of polydisperse hard spheres in space, determine the scattering cross sections with special care to avoid finite size effects of the simulation box, and then adjust the configurations such that the experimental scattering curve is reproduced, within error bars. First, the box size needs to be set in agreement with the characteristics of the scattering experiment. In practice, $L_{box}$ is set by the experimental minimum q-value, $L_{box} = 2\pi/q_{min}$. As the volume fraction of particles $\Phi$ is a key experimental parameter, the total number of particles N is fixed by $L_{box}$ and $\Phi$, as well as the average particle volume $V_P$ deduced from the experimental particle size distribution function measured in an independent form factor experiment. To fix ideas, in the experimental system described in this article, there are approximately 1200 particles in the simulation box for each percent of volume fraction, i.e., up to several thousand for the systems studied here. In all cases, the volume fractions were low enough such that an initial configuration could be generated easily by placing the beads with periodic boundary conditions in the box, and avoiding particle overlap. In order to study the impact of the initial configurations, we have generated more or less pre-aggregated configurations, and then checked the influence of pre-aggregation on the result. Pre-aggregation was parametrized by α, which is the fraction of NPs called 'seeds' that are distributed randomly in the beginning, without overlap, in the simulation box. In this context and throughout the article, the term 'random dispersion' is used to describe assemblies without correlations besides the ones caused by excluded volume. Then all following beads are put in contact with randomly chosen particles (seeds or not) that have already been positioned. $\alpha = 100\%$, e.g., gives a random distribution without any pre-aggregation, as every particle is a seed (but none is placed next to it on purpose), whereas the lower limit $\alpha = 1/N$ gives a single aggregate made of all particles around a single seed. The latter configuration is highly non homogeneous across the simulation box, and this undesired lack of homogeneity was found to become dominant for $\alpha < 1\%$. In the SI, $\alpha$ is shown to have only little impact on the final result, over a large range $1\% < \alpha < 100\%$, and we have set $\alpha$ to 5%.

After defining an initial configuration, spatial distributions of polydisperse hard spheres of scattering compatible with the measured intensities have been determined using a reverse Monte Carlo simulation combined with simulated annealing [70]. [55] Individual Monte Carlo steps are performed by randomly choosing a particle in the box, and moving it in a random direction, with step-length $\Delta$, and repeated on average once for each NP in the box, taking excluded volume into account at all times. This defines one simulation time step. In order to access configurations with close contact more quickly, half of the steps have been chosen to be random jumps towards contact with another randomly chosen bead. These steps are illustrated in Figure 1a. A classical metropolis algorithm was used to decide on the acceptance of each individual step, [71] involving an exponential weight of the 'energetic' cost $\Delta\chi^2$ of the move: $\exp(-\Delta\chi^2/\chi^2_{eff})$. The latter is defined as the increase in $\chi^2$ as given in the Methods section, and expresses the difference between experimental and simulated intensities. The effective 'temperature' of the process is given by $\chi^2_{eff}$, and it measures the acceptability of a move, which might worsen the agreement between the intensities. The idea of simulated annealing is then to decrease this 'temperature' until a series of particle configurations with acceptable $\chi^2$ is found. As the MC steps depend on $\Delta$, this last parameter is also decreased in order to fine-tune the final structures accordingly, following an exponential decay: $\Delta \approx \Delta_0 \gamma^n$, and similarly for $\chi^2_{eff}$, with $\gamma < 1$ (in practice: 0.98 and 0.905 for $\Delta$ and $\chi^2_{eff}$, respectively) and n being the number of time steps, and $\Delta_0$ some particle scale chosen to be the average radius, $<R> = 12.6$ nm. Both the imposed decrease of $\Delta$ and $\chi^2_{eff}$ are plotted in Figure 1b, together with the resulting MC success rate (defined by the sliding average fraction of allowed steps leading to a decrease or allowed small increase – by virtue of the MC-Boltzmann-criterion – of $\chi^2$, thus not taking collisions into account), and the decrease of $\chi^2$, as a function of the number of MC attempts to move each particle. The decrease is stopped once a satisfying agreement recognized by a small $\chi^2$ is found, and the simulation is continued with fixed parameters for averaging. During this averaging phase, multiple configurations compatible with the scattering are explored. This non-uniqueness of the configuration is a natural consequence of the ill-posed inverse scattering problem: there is much less information in the scattered intensity than in the particle configurations, and many configurations may correspond to the same scattering cross section, within error bars defined by the final $\chi^2_{eff}$. The average is thus performed over many such configurations.

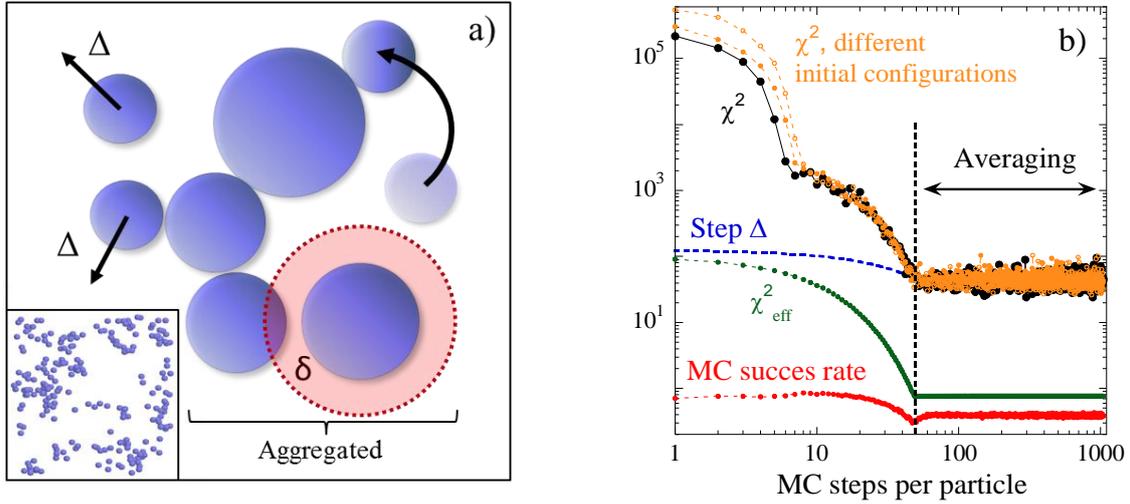

**Figure 1:** (a) Illustration of Monte Carlo moves (Δ, or jump to contact), aggregate recognition parametrized by δ, and a snapshot of a 70 nm-thick slice of the simulation box. (b) Example of simulated annealing as a function of the number of MC attempts to move each particle. $\chi^2$ expresses the quality of the fit, different initial configurations with α = 5% (orange) and 20% (black) are shown. Δ is the step-length of moves in Å, and $\chi^2_{eff}$ is the effective temperature. The MC success rate takes only $\chi^2$-decisions into account. Once $\chi^2$ crosses a critical value (after around 70 MC attempts), averaging starts.

The details of the decrease of the parameters Δ and $\chi^2_{eff}$ may be of importance for the structures selected by the algorithm. It is noted, however, that the final series of configurations is far from jammed (again, due to the low Φ), and that therefore many statistically equivalent configurations are reached. For comparison, $\chi^2$-values of different initial configurations have been superimposed in Figure 1b, two for α = 5%, and one for α = 20%. Note that the initial condition is defined by the parameter α, but also by the initialization of the random number generator used by the algorithm. The corresponding $\chi^2$-functions are seen to converge to about the same final $\chi^2$, i.e., the final fit quality is equally good.

The determination of $\chi^2$ for each particle configuration relies on a trustworthy calculation of the scattered intensity. In principle, there are several ways of calculating such intensities. For monodisperse systems, the pair correlation function is usually determined and Fourier-transformed. Polydisperse systems, however, require binning in discrete particle sizes which reduces the quality of the analysis. Staying in reciprocal space, this shortcoming can be avoided. There, a first method is based on the Debye isotropic averaging based on summing $\sin(rq_{ij})/rq_{ij}$-terms.[72] It is very efficient to calculate structure factors of aggregates,[55] but it is not appropriate for particle assemblies representing infinite systems, as it includes the low-q upturn to the total mass of the assembly, treated as a big aggregate, instead of smaller interacting objects. A second method – based on treating the system as the unit cell of a giant crystal – has been proposed by Frenkel in the 80s [73]. It has been applied to different systems, e.g., to polymers by Pedersen [74]. It is based on a calculation of the scattering amplitude in specific directions in space (usually 13), summing complex exponentials, followed by squaring. This method is very efficient, but potentially lacks statistics at high-q vectors due to the limited isotropic averaging. As shown below, it is straightforward to include polydispersity, where each particle j is described by the Fourier transform of its scattering length density Δρ. The latter transform is called the form factor amplitude, which – normalized to 1 at low q – reads $F_j(q)$. This function multiplied by the particle volume $V_j$ and contrast Δρ, and squared, is commonly called the form factor $P_j(q)$, and it can be averaged over all particles, giving $\bar{P}(q)$. Polydispersity can be included in the Frenkel formalism by multiplying the phase exponential by the normed form factor amplitude, contrast Δρ, and the particle volume $V_j$. The intensity scattered by N polydisperse particles located at positions $(x_j, y_j, z_j)$ then reads:

$$I(q_p) = \left| \sum_{j=1}^{N} \Delta\rho \, F_j(q_p) V_j \exp\left[-ip2\pi \frac{hx_j + ky_j + lz_j}{L_{box}}\right] \right|^2 \quad (3a)$$

$$S(q_p) = \frac{I(q_p)}{\bar{P}(q_p)} \quad (3b)$$

where the norm of the scattering vector is given by $q_P = p \, 2\pi \frac{\sqrt{h^2+k^2+l^2}}{L_{box}}$, with p = 1, 2, 3… The experimental structure factor $S(q_p)$ with its apparent isothermal compressibility is obtained by dividing by the average form factor $\bar{P}(q_p)$ in eq (3b). Note that the contrast term Δρ is supposed to be homogeneous for all spheres here, and it thus cancels in this ratio. Configurational averages are then performed on $S(q_p)$, simply written S(q) in what follows.

By continuing the random particle displacements under the condition of agreement with the experimental intensity defined by the Boltzmann factor, a sequence of statistically equivalent particle configurations is obtained. We have checked that these particle configurations are statistically independent, based on the time autocorrelation function of the low-q intensity (see SI). The intensities can then be averaged to obtain smooth intensities, and configurational averages. Moreover, all configurations can be analyzed, as done below for the state of aggregation. Alternatively, single snapshots of configurations in the box can be generated for illustration, and an example of a slice is included in Figure 1a.

In order to access the average state of aggregation corresponding to the scattered intensities, or any other property of interest, an aggregate recognition algorithm can be run on the particle configurations. Its result is the mean number fraction of aggregates of a given aggregation number $N_{agg}$ present during the averaging procedure (Figure 1b) in a simulation box of scattering compatible with the experimental intensity. The heart of this algorithm is to recognize if two given particles have a surface-to-surface distance lower than some critical distance δ, as illustrated in Figure 1a. Note that particles may be in touch across the periodic boundary conditions. The choice of this parameter δ needs to be discussed. One would intuitively expect that δ = 0 might be the appropriate choice. Small-angle scattering being a low-resolution method, it is, however, impossible to distinguish particles close-by from effectively touching particles. Accordingly, a non-zero δ allows counting particles as 'in contact' even if they are at a small distance in the simulation. The natural length scale of the problem is the particle radius, and we have set δ = <R>. Note that the effective volume fraction of the particle plus a layer of thickness δ/2 increases with the effective radius to the third power, and choosing δ = 2<R> would result in an 8 times higher volume fraction, virtually aggregating the entire simulation box for experimentally relevant Φ. The choice of δ is thus rather restricted to values around <R>. It may be noted that this difficulty exists also in the analysis of TEM-pictures, where it is unclear due to the projection in 2D of a three-dimensional slice if two particles apparently in touch are actually in contact.

**Aggregation of surface-modified NPs.** We have performed surface modification of silica nanoparticles ($R_0$ = 12.5 nm, σ = 0.12) in suspension with octyl-silanes of different functional groups as indicated in Table 1. In Figure 2a, the scattered intensities of surface-modified nanoparticles at 1%v suspended in the ethanol-water reaction mixture (63/37 by vol) are compared to the one of the bare NPs in the same solvent. From the increase in the low-q intensity with respect to the bare NPs, one may immediately deduce that the surface modification was successful. Intuitively, the low-q scattering seems to indicate that the standard triethoxy silane forms the biggest aggregates, whereas the trimethoxy version displays lower scattering and thus average aggregation, and the monomethoxy function $C_{8mm}$ the lowest one. Only the bare NPs seem to be almost perfectly dispersed in this solvent. The quantitative analysis of the 3D-structure in terms of aggregation by RMC will be shown to be consistent with the above picture (i.e., aggregation $C_8$ > $C_{8m}$ > $C_{8mm}$ > bare). The simple analysis probably works due to the absence of any obvious manifestation of varying inter-aggregate interactions for the different grafts, and namely of peaks in the intermediate-q range. Intensity modifications through weak low-q depressions or enhancements induced by interactions, however, cannot be ruled out a priori, and a method capable separating these interactions from the aggregation is needed.

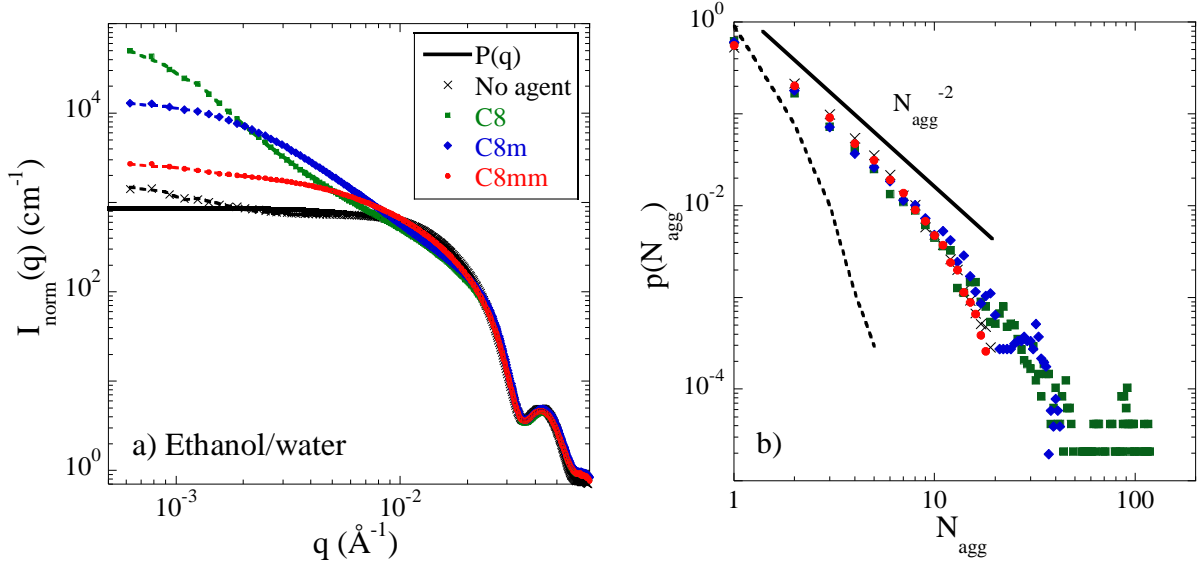

**Figure 2:** (a) SAXS scattered intensities normed to NP form factor in water (continuous line, $R_0 = 12.5$ nm, $\sigma = 0.12$) of bare and surface-modified ($C_8$, $C_{8m}$, $C_{8mm}$) NPs in hydro-alcoholic suspension (63%v ethanol, $\Phi = 1\%$v). Dotted lines are the average RMC fits. (b) Aggregate mass distributions of the same samples. The distribution for a random dispersion of the same NPs with excluded volume ($\Phi = 1\%$v, $\delta = <R>$) is represented by a dotted line.

The particle configurations generated by the RMC algorithm during the averaging phase – see Figure 1b – may be used to extract the quantities of interest for the present study. Each series of symbols in Figure 2a is superimposed to a dotted line representing the average RMC fit of the intensity. Note that the agreement is better than the size of the symbols. In Figure 2b, the resulting aggregate mass distributions corresponding to the intensities are plotted, using $\delta = <R>$. These distribution functions represent the real-space analogues of the intensities, without however the inter-aggregate interactions. In other words, we have chosen to focus on the mass of the aggregates, by setting up an analysis which allows ignoring where aggregates are positioned with respect to one another. This amounts to having taken out all partial inter-aggregate structure factors off the scattered intensity, a result which would not have been possible directly in reciprocal space. This feature will be of particular importance for higher volume fractions as encountered in the nanocomposites, where interactions are more dominant. The aggregate distribution functions in Figure 2b decay strongly for all surface modifications, following roughly a $N_{agg}^{-2}$ power law. The distributions are mostly identical for small aggregates, and the differences become important only in the large $N_{agg}$-range. Above $N_{agg} \approx 10$, aggregates become very rare. Bare NPs and $C_{8mm}$-modified ones do not have aggregates above $N_{agg} \approx 20$, while $C_{8m}$-modified aggregates extend up to $N_{agg} \approx 40$, and $C_8$-modification even above 100. For comparison, we have calculated the aggregate mass distribution function for the same parameters ($\Phi = 1\%$v, $\delta = <R>$) for randomly distributed polydisperse spheres with excluded volume of the same size. The result is superimposed to the experimental distributions in Figure 2b. Clearly, the structure of the suspension is not random, and considerably more aggregated than the random distribution, which tends to zero with a high power law exponent approaching minus 4. In hydro-alcoholic suspension, $C_8$-surface modification thus favors the formation of bigger (but still finite) aggregates, whereas all others limit aggregation even further. Counterbalancing electrostatic repulsion, the hydrophobic interaction between particles is thus strong enough to induce aggregation up to a maximum of ca. one hundred particles. As opposed to the monofunctional grafts, it is also possible that polycondensed groups form on the NP surface in particular in presence of water. [75, 76] Such hydrophobic patches may induce additional attractive interactions between NPs in the still hydrophilic solvent, in analogy with surfactant micelles or proteins which may adsorb on NP surfaces [77-83] and destabilize the colloidal suspension [65-67, 84-86]. $C_{8m}$ represents an intermediate case, where the trifunctional grafting functions may still lead to the formation of polycondensed patches on the silica surface, favoring attractive interactions, with a slightly different behavior which can only be attributed to the different reactivity of the methoxy functions. The observed lowest aggregation with $C_{8mm}$-grafting, finally, must be due to the different reaction mechanism of mono-functionalized molecules, in spite of the identical octyl-chain of the coating agent. With $C_{8mm}$, there are no other lateral groups that can react with neighboring molecules, and patches cannot form. Moreover, the final NP hydrophilicity is affected via the number of remaining silanol groups after grafting. Monofunctional grafts interact with only one silanol, leaving unreacted silanols to maintain the electrostatic contribution to the dispersion in hydroalcoholic suspension.

In the next step, the suspensions of surface-modified NPs have been dialyzed into MEK (see Methods section). Changing the solvent by dialysis is expected to have a considerable impact on the dispersion of the surface-modified NPs, due to the change in polarity, as well as solubility of the alkyl chains. Indeed, MEK is still quite polar ($\varepsilon_{MEK}$ = 18.5, to be compared to the water-ethanol mixture, between 24.5 and 80 for the pure solvents, respectively), and it is an excellent solvent of the alkyl chains. In Figure 3a, the scattering of the $C_8$-surface-modified NPs is plotted, displaying lower intensities and thus better dispersion than in ethanol-water. The type of grafting function of the silane seems to have a strong influence on the NP dispersion, with an inverted order in this solvent with respect to the previous one (Figure 2a): in MEK, the triethoxysilane $C_8$ seems to have the best dispersion, presenting a repulsive interaction peak at intermediate q, followed by the $C_{8m}$ (which is thus again in the middle of the three, and also shows some structure at intermediate q), and finally the monomethoxysilane $C_{8mm}$ has the highest low-q scattering.

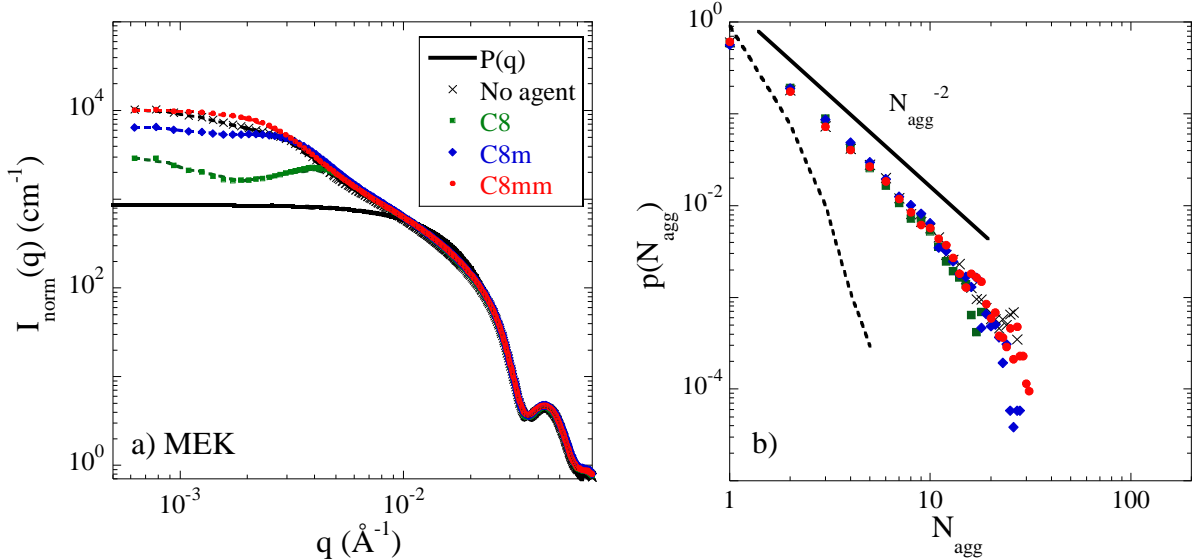

**Figure 3:** (a) SAXS scattered intensities normed to NP form factor in water (continuous line, $R_0$ = 12.5 nm, $\sigma$ = 0.12) of bare and surface-modified NPs ($C_8$, $C_{8m}$, $C_{8mm}$) in MEK ($\Phi$ = 1%v). Dotted lines are the average RMC fits. (b) Aggregate mass distributions of the same samples. The distribution for a random dispersion of the same NPs with excluded volume ($\Phi$ = 1%v, $\delta$ = <R>) is represented by a dotted line.

As with the ethanol-water mixture, the combination of remaining electrostatic repulsion with the solvophilic character of the surface modification needs to be invoked to provide an explanation for the observed order in the scattered intensities. In MEK, electrostatic interaction should still be strongest for bare and monofunctionalized NPs, but globally weaker than in ethanol-water. As one can see with the aggregation of the bare NPs, this effect is not sufficient to guarantee individual dispersion in this solvent. In presence of grafts, the resulting order in aggregation from best dispersion for $C_8$-modification, followed by $C_{8m}$, and $C_{8mm}$ can be rationalized through the additional steric stabilization caused by solvated grafted molecules. Steric repulsion is observed to be strongest for the trifunctional molecules. These molecules form a solvated and thus repulsive layer on the NPs, and may possibly also form solvated patches. This contribution to the repulsion becomes weaker between monofunctionalized NPs, which is apparently compensated by the stronger electrostatic repulsion due to the preservation of surface silanols. As a result, these NPs thus display a similar behavior as the bare NPs. In Figure 3a, a short fractal regime may be present, in particular for the $C_{8mm}$-sample, corresponding to a fractal dimension of 1.9. It extends up to the Guinier regime of the form factor ($\approx$ 1.5 $10^{-2}$ Å$^{-1}$), i.e. it corresponds to a linear aggregate dimension of no more than 3 or 4 NPs. It is arduous to conclude on fractality for such small particle numbers. It is difficult to analyze quantitatively the 3D-structure corresponding to this family of scattered intensities, due to the superposition of aggregate formation as indicated by the increase of the low-q structure, polydispersity in aggregate size, and aggregate interaction. The latter may add peaks, or possibly a low-q depression or enhancement (as with $C_8$-NPs in Figure 3a). The RMC analysis will show that although the interactions between aggregates are different for the different grafts in MEK, the aggregates themselves are quite comparable, and rather small.

The dotted lines in Figure 3a represent the RMC fits of the MEK-samples, which are again not distinguishable from the symbols representing the experimental data points. Moreover, the suspension structure is again very different from the random structure processed for comparison in an identical way. After application of the aggregate recognition algorithm, the corresponding average aggregate mass distributions are plotted in Figure 3b (see SI for a few pictures of aggregates). Given the differences in the scattering, the mass distribution functions in MEK superimpose surprisingly well. Above some $N_{agg} = 20 - 30$ particles, there are no more big aggregates, for any surface modification. Below this value, the distribution function follows again approximately a $N_{agg}^{-2}$ power law, as indicated in the Figure. The combined RMC and aggregate recognition approach thus allows to deliberately ignore the inter-aggregate interactions in this solvent, and highlight the aggregate mass distribution. The latter function is found not to depend strongly on the presence or type of surface modification, and goes to zero quickly. Apparently, MEK is a rather good solvent both for bare particles, and the grafted NPs studied here, possibly for a combination of effects, namely both steric and electrostatic stabilization. One has to go in great detail at the highest aggregation numbers, in order to find some bigger aggregates for $C_{8mm}$, which remain extremely rare. The presence of a well-defined scattering peak in Figure 3a for the $C_8$-sample is intriguing when one considers the wide polydispersity in aggregate mass as given in Figure 3b. Our explanation is that the assembly of all aggregates is nonetheless structured on the scale of $1/q_{peak}$. Real-space pictures of simulations shown in the SI seem to confirm this hypothesis.

The particle suspensions in MEK are precursor solutions for the formation of nanocomposites. The latter are produced by mixing with the appropriate amounts of styrene-butadiene chains, followed by evaporation of the solvent. The final silica volume fractions in the PNCs were determined by TGA, and lie between 1% and ca. 3%, i.e., nanocomposites are about as dilute as the precursor suspensions (1%v). One of the key questions of this commonly used solvent casting technique is if the structure in suspension predetermines the NP dispersion in the polymer matrix. By comparing the SAXS intensities shown in Figure 4a to the ones in Figure 3a, it is observed that the intensities remain similar in order, but are globally increased in magnitude. Among the curves, the intensity of $C_{8mm}$ is still the highest one, indicating highest aggregation, whereas the intensity of $C_{8m}$ is almost unchanged, and the one of $C_8$-modified NPs loses its peak which is transformed into a much weaker structural feature, a shoulder.

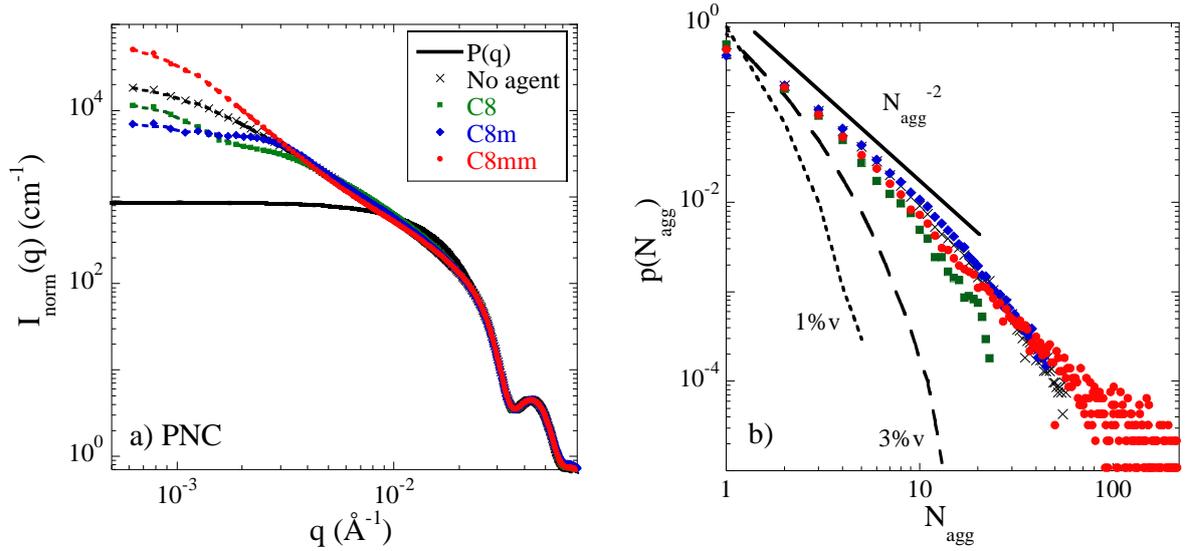

**Figure 4: (a)** SAXS scattered intensities normed to NP form factor in water (continuous line, $R_0 = 12.5$ nm, $\sigma = 0.12$) of bare (2.6%v) and surface-modified NPs ($C_8$, $\Phi = 1.2$%v; $C_{8m}$ 3.2%v; $C_{8mm}$ 2.8%v) in PNCs. Dotted lines are the average RMC fits. **(b)** Aggregate mass distributions of the same samples. Distributions for a random dispersion of the same NPs with excluded volume ($\delta = \langle R \rangle$) are represented by a dotted ($\Phi = 1$%v) and a dashed line ($\Phi = 3$%v).

It is concluded that aggregates in nanocomposites are probably bigger in mass than the ones in the MEK-precursor solution, and that interactions are modified. However, due to the occurrence of different large-scale organization in space leading to different interactions and thus average structure factors, which moreover are affected by changes in concentration, it is very difficult to extract quantitative information from these curves. Applying the RMC simulation with aggregate recognition gives the aggregate mass distribution functions – regardless of inter-aggregate interactions – shown in Figure 4b. These distributions are quite different for the different surface-modifications, and again much more aggregated than the random dispersion. The $C_8$-modification leads to the best dispersion, with a mass distribution falling quickly below the $N_{agg}^{-2}$ power law. The other three samples follow this power law quite closely, but extend differently into the high-$N_{agg}$ regime. In particular, $C_{8mm}$ is found to favor the formation of big aggregates, counting up to 200 nanoparticles. This might be due to the enhancement of the destabilization during drying and concentration induced in the casting phase.

In order to further analyze the mass distribution functions, one may calculate simple observables characterizing their average and the width. The first moment gives the average, $\langle N_{agg} \rangle$, and the second moment $\langle N_{agg}^2 \rangle$ may be normalized by $\langle N_{agg} \rangle$, then giving an indication of the width of the distribution by weighting higher masses more strongly. This quantity should also describe the low-angle scattering in absence of interaction, which is proportional to $\langle V^2 \rangle / \langle V \rangle$, where V is the volume of objects correlated in space, i.e., aggregates, and thus reduces to $\langle N_{agg}^2 \rangle / \langle N_{agg} \rangle$, at least in first order, i.e., neglecting the (weak) particle polydispersity. Moreover, the present determination of aggregate distribution functions has a sharp cut-off parameter $\delta$, and the $\langle N_{agg}^2 \rangle / \langle N_{agg} \rangle$ are thus expected to be smaller than experimental low-q intensities in absence of interaction. As we will see, the order of magnitude, and more importantly, the order of the intensities are correctly described. Note that, for the sake of completeness, the evolution of the polydispersity index $<N_{agg}^2>/<N_{agg}>^2$ is given in SI.

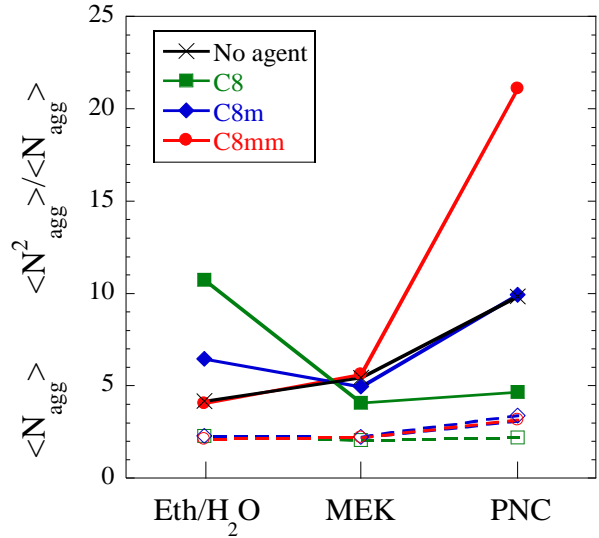

**Figure 5:** First moment $<N_{agg}>$ (empty symbols) and width-parameter $<N_{agg}^2>/<N_{agg}>$ (plain symbols) of the mass distribution functions in the three suspension media (hydro-alcoholic mixture, MEK, and polymer).

In Figure 5, the parameter characterizing the width of the mass distribution functions $<N_{agg}^2>/<N_{agg}>$ extracted from the data in Figures 2b, 3b and 4b, is plotted for the three suspension media. As the width is mainly related to the existence of large aggregates, the evolution observed in Figure 5 summarizes the previous discussion of the distribution laws: in the ethanol-water mixture, $C_8$-surface modification forms the biggest aggregates; in MEK, all aggregates are rather small, with $C_{8mm}$ favoring aggregation as much as the bare NPs, due to the combination of steric and electrostatic repulsion; in polymer nanocomposites, finally, $C_{8mm}$ induces the worst dispersion and $C_8$ the best, suggesting that the tendency of the NPs to aggregate in MEK is enhanced when adding the polymer. Coming back to the first moment, $<N_{agg}>$ evolves only slowly with surface modification and changes in the suspension medium, although some increase from MEK to PNCs is also visible. This is due to the predominance of aggregates with low aggregation numbers in all samples, leading to overlapping functions at low $N_{agg}$ in Figures 2b, 3b, and 4b. A set of representative TEM pictures is given in the SI, together with a direct comparison with a slice of the simulation box.

The results shown in Figure 5 may be used to discuss a possible transfer of the quality of the dispersion from MEK into the polymer nanocomposites. It is observed that the order of the width of the mass distributions is conserved from MEK to the PNCs. On the other hand, only the $C_8$-modified

NPs keep a comparable mass distribution, whereas the others become wider, in particular the $C_{8mm}$ one, which shows a strong evolution towards bigger aggregates in the polymer matrix. This suggests that if aggregates are stable in MEK (as with $C_8$), then the addition of polymer is sufficient to 'freeze' the state of aggregation during the drying process, which is characterized by a strong increase of viscosity. The most unstable aggregates, made of $C_{8mm}$-grafted NPs, however, seem to be further destabilized in the polymer solution, and aggregate before being frozen in by the drying process.

**Conclusions**

In the present paper, we have proposed a combined reverse Monte Carlo and aggregate recognition analysis for structural investigations by SAXS of dispersions of polydisperse nanoparticles suspended in any medium. The procedure has been applied to nanoparticles dispersed in solvents and in polymer nanocomposites, with various states of aggregation triggered by surface modifications of the NPs with silane molecules carrying different grafting functions. The sequence of the media corresponds to the reaction pathway from bare suspended NPs in water, to ethanol-water mixtures allowing the grafting reaction, phase transfer into MEK to prepare solvent casting with polymer molecules, and finally the formation of polymer nanocomposites. Our analysis of the experimental SAXS data allows to follow aggregation in these different media, via the distribution functions of aggregate mass, independently of any possible interactions between the aggregates in the medium. This is achieved by analyzing real space particle configurations obtained by reverse Monte Carlo, a feature which shall be particularly useful for samples at high NP content, as typically encountered in polymer nanocomposites.

The aggregate mass distribution functions have been determined for all bare and surface-modified NPs in hydro-alcoholic mixture, in MEK, and in nanocomposites. It has been found that surface-modified particles are mostly not well dispersed in ethanol-water mixtures, due to the hydrophobicity of the grafts, unless some electrostatic repulsion remains after grafting. The grafting reaction affects the number of silanol groups still available on the NPs, and a higher number of unreacted silanols – as encountered for bare and $C_{8mm}$-nanoparticles – induces more repulsive electrostatic interactions between particles and aggregates. This is possible as long as the medium is polar enough, which is the case for both alcohol and MEK. In MEK, NPs are globally better dispersed, due to a combination of remaining electrostatic and steric repulsion caused by a solvated grafted layer, forming only small NP aggregates below $N_{agg} = 20$. Our analysis shows that the mass distributions are similar for the different grafts in MEK, although the various interactions make the scattered intensities look different. The impact of the grafts, either attractive or sterically repulsive, depends on the solvent, and it seems to be enhanced due to the presence of patches formed by polycondensation of the trimethoxy- or triethoxysilanes on the silica surface. In polymer nanocomposites, finally, the aggregation is enhanced, presumably due to the drying phase where particle concentration and thus interaction increases naturally. The order of the curves being maintained from the MEK suspension to the PNCs, this suggests that there is at least a partial transfer of the pre-aggregated state of nanoparticles in suspension into the polymer matrix.

We have commented that a short possibly fractal domain may exist in some of the scattering curves. As aggregates are rather small, this is difficult to confirm. It is true, however, that our reverse Monte Carlo algorithm naturally favors the most disordered structures, and in particular the most polydisperse assembly of aggregates. This corresponds to the automatic minimization of the information content in our analysis, by choosing the more probable structures. Any more ordered structure, like monodisperse fractals, will thus not be retrieved. Another example of a more ordered structure, linear aggregates, is explicitly discussed in the SI. Our analysis shows that scattering from such a partially ordered structure is interpreted as a polydisperse assembly of comparable aggregation number, details of which are still to be investigated. On the other hand, the agreement between the statistical indicators and the low-q intensity shows that relevant aggregate information is recovered by RMC, regardless of interaction.

A possible perspective of our work is to analyze the impact of increasing volume fractions on aggregate mass distributions, i.e., in a situation where inter-aggregate structure factors are known to be dominant. The signature of NP percolation, e.g., should be the recognition of a single giant aggregate filling the entire simulation box. This RMC-approach will be applied to a nanocomposite system with silanes of various alkyl lengths in a future paper. Alternatively, one might also investigate the reorganization of nanoparticles in dense assemblies or nanocomposites, including aging phenomena, under strain [87] or annealing at high temperature. This might open the road to detailed studies of, e.g., the Payne effect, [88, 89] where the mechanical response of a nanocomposite system is found to evolve in a non-linear way with strain, presumably due to particle reorganizations inducing changes in the aggregate mass distribution.

**Acknowledgements.** The authors are thankful for support by the ANR NANODYN project, grant ANR-14-CE22-0001-01 of the French Agence Nationale de la Recherche. Help from Ty Phou (L2C) with formulation is gratefully acknowledged. We are indebted to Jan Skov Pedersen for reminding us of the properties of equation 3.

**FOR TABLE OF CONTENTS USE ONLY**

Aggregate formation of surface-modified nanoparticles in solvents and polymer nanocomposites

Dafne Musino, Anne-Caroline Genix, Thomas Chaussée, Laurent Guy, Natalia Meissner, Radoslaw Kozak, Thomas Bizien, Julian Oberdisse

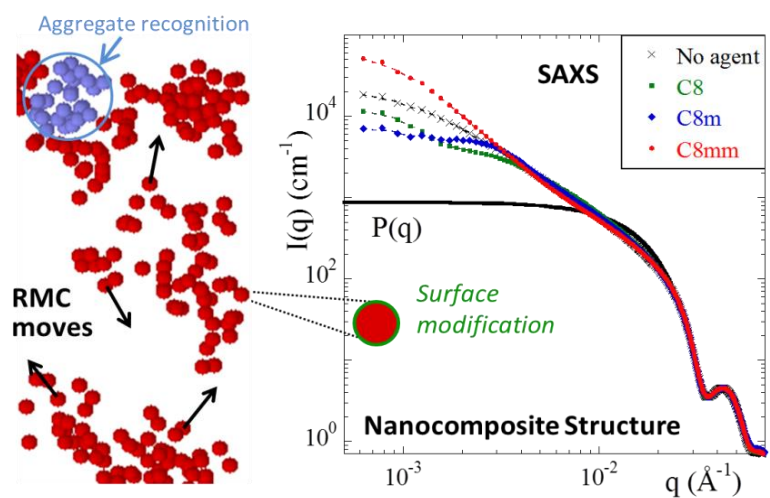